\documentclass[
twocolumn,
showpacs,preprintnumbers,
bibnotes,
 amsmath,amssymb,
 aps,
 pra,
superscriptaddress,
longbibliography,
]{revtex4-1}

\usepackage[version=3]{mhchem} 
\usepackage[dvipsnames]{xcolor}
\usepackage{textcomp}
\usepackage{graphicx}
\usepackage{amsmath}
\usepackage{fixmath}
\usepackage{amsfonts}
\usepackage{amssymb}
\usepackage{siunitx}

\newcommand{\micron}{\hbox{\textmu}\text{m}}

\graphicspath{ {./figures/} }

\begin{document}

\title{Nonlinear self-action of ultrashort guided exciton-polariton pulses \\
in dielectric slab coupled to 2D semiconductor} 

\author{F.A.~Benimetskiy}

 \affiliation {School of Physics and Engineering, ITMO University, St. Petersburg, Russia}
\author{A.~Yulin}
 \affiliation {School of Physics and Engineering, ITMO University, St. Petersburg, Russia}
\author{A.O.~Mikhin}
 \affiliation {School of Physics and Engineering, ITMO University, St. Petersburg, Russia}
\author{V.~Kravtsov}
 \affiliation {School of Physics and Engineering, ITMO University, St. Petersburg, Russia}
 \author{I.~Iorsh}
 \affiliation {School of Physics and Engineering, ITMO University, St. Petersburg, Russia}
\author{M.S.~Skolnick}
\affiliation
{Department of Physics and Astronomy, University of Sheffield, Sheffield S3 7RH, UK}
\author{I.A.~Shelykh}

  \affiliation {School of Physics and Engineering, ITMO University, St. Petersburg, Russia}
  \affiliation {Science Institute, University of Iceland, Dunhagi-3, IS-107, Reykjavik, Iceland}
\author{D.N. Krizhanovskii}
\affiliation
{Department of Physics and Astronomy, University of Sheffield, Sheffield S3 7RH, UK}
\author{A.~Samusev}
 \affiliation {School of Physics and Engineering, ITMO University, St. Petersburg, Russia}

\begin{abstract}
Recently reported large values of exciton-polariton nonlinearity of transition metal dichalcogenide (TMD) monolayers coupled to optically resonant structures approach the values characteristic for GaAs-based systems in the regime of strong light-matter coupling. Contrary to the latter, TMD-based polaritonic devices remain operational at ambient conditions and therefore have greater potential for  practical nanophotonic applications.
Here we present the study of the nonlinear properties of Ta$_{2}$O$_{5}$ slab waveguide coupled to a WSe$_{2}$ monolayer. We first confirm that the hybridization between waveguide photon mode and a 2D semiconductor exciton resonance gives rise to the formation of exciton-polaritons with Rabi splitting of 36~meV. By measuring transmission of ultrashort optical pulses through this TMD-based polaritonic waveguide, we demonstrate for the first time the strong nonlinear dependence of the output spectrum on the input pulse energy. Our theoretical model provides semi-quantitative agreement with experiment and gives insights into the dominating microscopic processes which determine the nonlinear pulse self-action: Coulomb inter-particle interaction and scattering to incoherent excitonic reservoir. We also confirm that at intermediate pump energies the system supports quasi-stationary solitonic regime of pulse propagation. Our results are essential for the development of nonlinear on-chip polaritonic devices based of 2D semiconductors.
\end{abstract}

\maketitle


\section{Introduction}

Two-dimensional transition metal dichalcogenides (TMDs) are actively used as alternative materials to III-V semiconductors~\cite{weisbuch1992observation, kasprzak2006bose,amo2009superfluidity,amo2011polariton,lagoudakis2008quantized,stevenson2000continuous} to realise a strong light-matter coupling regime between excitons and photons in hybrid systems~\cite{schneider2018two}. Contrary to their bulk counterparts, monolayer TMDs are direct band-gap semiconductors possessing strong excitonic response in the visible diapason. The Wannier–Mott excitons in these materials are characterised by large binding energies (up to 200-500 meV for neutral excitons and $\sim$30 meV for charged excitons -- trions) and large exciton oscillator strengths~\cite{chernikov2014exciton,li2014measurement,wang2018colloquium}. This makes possible the formation of hybrid light-matter quasiparticles -- exciton-polaritons at cryogenic~\cite{dufferwiel2015exciton, barachati2018interacting, emmanuele2020highly, kravtsov2020nonlinear} and room temperatures~\cite{zhang2018photonic,cuadra2018observation, shan2021spatial,wurdack2021motional} in TMD-based systems with excitons resonantly coupled to a spatially onfined photonic mode. 

The main interest of exciton-polaritons is their hybrid nature  ~\cite{agranovich1957influence, hopfield1958theory}. Due to the photonic component, the effective mass of these quasiparticles is extremely small compared to masses of pure excitons, and their coherence length lies in the micrometer range \cite{Ballarini2017}. In the same time, the excitonic component gives rise to the efficient polariton-polariton interactions, which lead to strong nonlinear optical response \cite{Estrecho2019}. These remarkable properties make systems based on exciton-polaritons excellent candidates for practical realization of active all-optical on-chip photonic devices.

Recently, the nonlinear optical response has been investigated in TMD-based polariton systems where atomically-thin semiconductors were integrated with photonic structures such as open cavity~\cite{emmanuele2020highly}, dielectric Bragg stack supporting Bloch surface waves~\cite{barachati2018interacting}, and photonic crystal slabs supporting either leaky modes~\cite{zhang2018photonic} or bound states in continuum~\cite{kravtsov2020nonlinear}. Furthermore, the enhanced nonlinear response of polaritons in TMD-based structures has been shown for Rydberg excitons~\cite{gu2021enhanced}, trions~\cite{emmanuele2020highly} and indirect excitons in Moire-type structures \cite{Zhang2021}. The nonlinear polaritonic response typically manifests itself as a power dependent shift of polariton branches due to either blueshift of the exciton energy arising from the Coulomb interaction~\cite{Shahnazaryan2017} or the nonlinear quenching of Rabi splitting due to the phase space-filling effects~\cite{emmanuele2020highly}. Both mechanisms come into play with the increase of polariton density.

The magnitude of the nonlinear optical effects observed in polaritonic systems based on 2D semiconductors is comparable to 
that in quantum-dimensional devices based on III-V semiconductors~\cite{dufferwiel2015exciton, barachati2018interacting, emmanuele2020highly, kravtsov2020nonlinear, gu2021enhanced}. 
Despite the fact that a number of dynamic nonlinear effects such as solitons~\cite{walker2019spatiotemporal,walker2015ultra}, self-phase modulation~\cite{walker2019spatiotemporal}, parametric scattering~\cite{suarez2020demonstration}, lasing~\cite{Suarez-Forero:20} have been revealed in planar polariton waveguides based on GaAs, GaN and others, to the best of our knowledge none of them have been demonstrated in TMD-based polaritonic systems. 

\begin{figure*}[ht!]
\centering
\includegraphics[width=\textwidth]{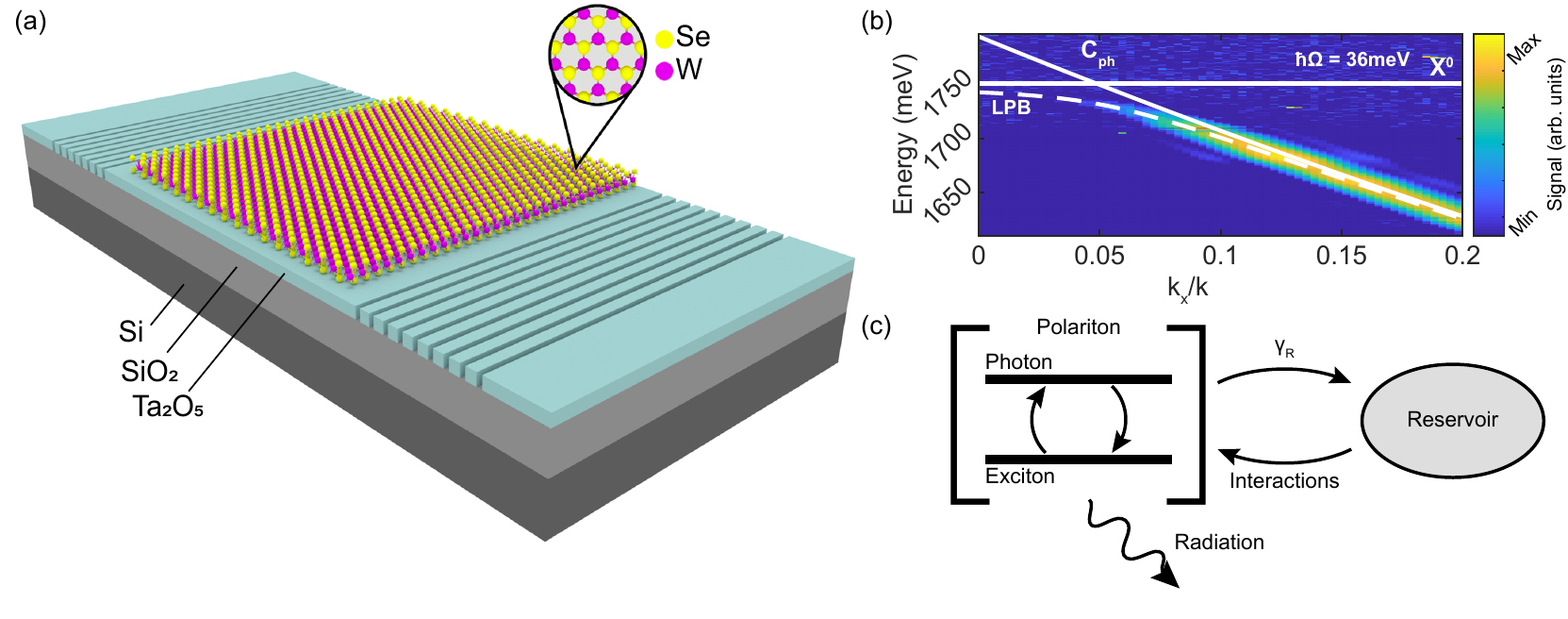}
\caption{\label{fig:1} (a) Sketch of a Ta$_2$O$_5$ slab waveguide integrated with a WSe$_2$ monolayer. Periodic gratings are used to couple and decouple femtosecond pulses. (b)  Angle- and frequency-resolved photoluminescence measured from the are of the decoupler covered by a TMD monolayer. Solid lines represent exciton ($X^0$) and photon ($C_{ph}$) modes. Dashed line shows the lower polariton branch disperion obtained from the fitting of the data with coupled oscillator model. (c)  Scheme showing the several processes accounted for in the model: exciton-photon coupling, scattering into the exciton reservoir, photon emission.}
\end{figure*}

In this work, we comprehensively study the nonlinear evolution of intense femtosecond polaritonic pulses propagating along a single-mode planar waveguide integrated with a WSe$_2$ monolayer. We first experimentally confirm the strong light–matter coupling between photonic mode of the waveguide and excitonic resonance in TMD. Further, we perform I-scan optical spectroscopy experiments revealing complex nonlinear reshaping of transmitted pulse spectrum depending on the energy of an input 
pulse. We then use a theoretical model allowing to reproduce the experimental results in a semi-quantitative manner. This becomes possible if both polariton nonlinearity driven by Coulomb exciton-exciton interaction and scattering of coherent excitons into incoherent reservoir are accounted for. Finally, relying on the model we analyze the spatio-temporal evolution of the ultrashort polariton pulses realized in the experiment. In particular, we confirm that the spectral broadening of the output signal experimentally observed at intermediate pump levels corresponds to the realization of quazi-stationary solitonic regime of pulse propagation. 


\section{Results}
\subsection{Experiment}
The studied sample is sketched in Figure~\ref{fig:1}(a) and consists of an unpatterned Ta$_2$O$_5$ slab waveguide coupled to WSe$_2$ monolayer surrounded by two coupler/decoupler gratings. In the first stage, the gratings 
were etched in a 90-nm-thick Ta$_2$O$_5$ layer deposited on SiO$_2$/Si (\SI{1}{\micro\metre}/\SI{525}{\micro\metre}) substrate. 
Further on, a mechanically exfoliated 50-$\mu m$-long monolayer of  WSe$_2$ was transferred on the surface of a slab waveguide between gratings. Such a configuration allows us to study nonlinear processes that occur during the propagation of laser pulses through the structure. Figure~\ref{fig:1}(b) shows the measured lower polariton dispersion with the avoided crossing between uncoupled photon and exciton modes characterized by a Rabi splitting of $\hbar\Omega_R~\sim~36\pm1$~meV. 

\begin{figure}[h]
\includegraphics{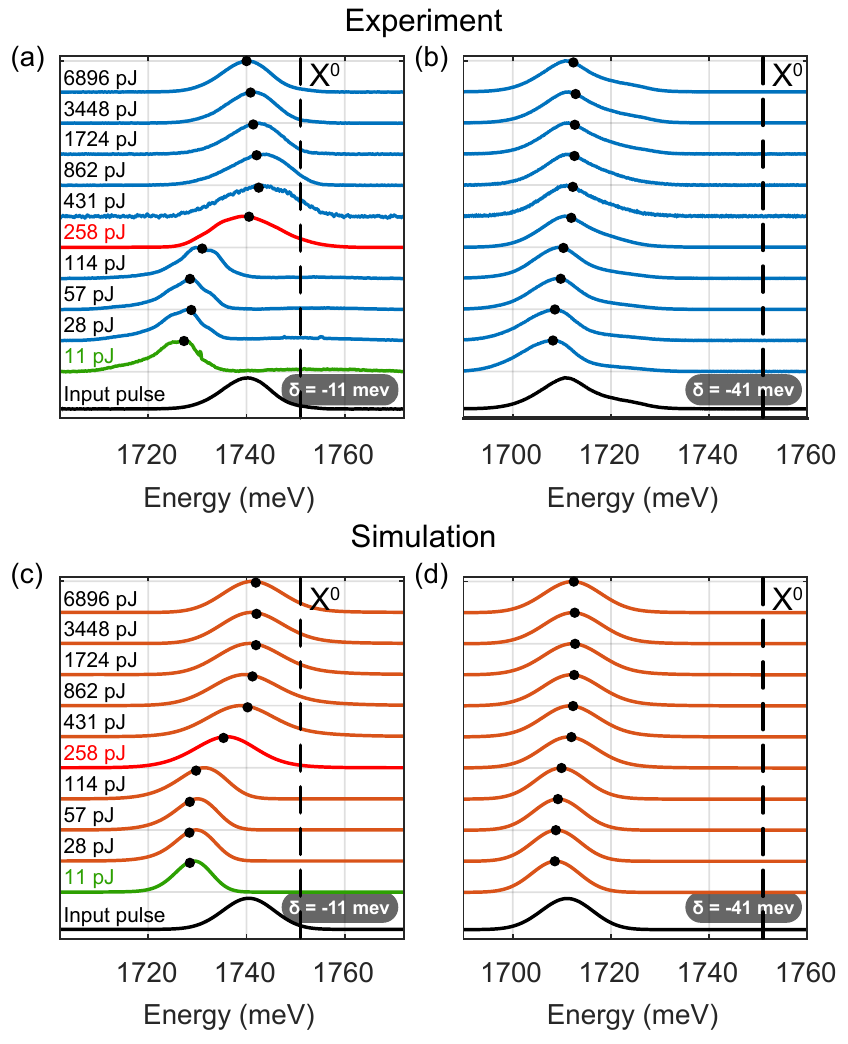}
\caption{\label{fig:2} (a-b) Measured normalized transmitted pulse spectra for the input pulse energy increasing from the bottom to the top. The estimated input pulse energies are specified above the respective curves. The input pulse spectrum is shown at the bottom of each panel. Its shape is assumed to match the transmitted pulse spectrum under high-energy excitation (the details are given in the text). The dashed vertical line (X$^0$) denotes the frequency of the uncoupled exciton resonance. Presented data correspond to negative detunings of the input pulse central frequency of (a) $-11~meV$ and (b) $-41~meV$ from the exciton resonance. (c-d) Results of the theoretical modelling, corresponding to the upper panels. Depending on the detuning and power of the input pulse, the output spectrum demonstrates complex behavior of both its linewidth and spectral position. }
\end{figure}

In the experiment, the sample was placed into a closed-cycle helium cryostat integrated with a setup for the Fourier plane imaging allowing to perform the angle-resolved spectroscopy measurements at 7K. The sample is excited by laser pulses with a square spectral envelope of 19~meV corresponding to a $\sim$130~fs temporal full width at half maximum (FWHM). The pulses are coupled at an optimized angle through the input grating, propagate
\SI{100}{\micro\metre}
along the waveguide, and decouple through the output grating. Note that the effective length of the interaction between the excitons and the waveguide mode is $\approx$\SI{50}{\micro\metre}, which is defined by the size of the area covered by a WSe$_2$ monolayer along the propagation direction.

\begin{figure}[!h]
\includegraphics{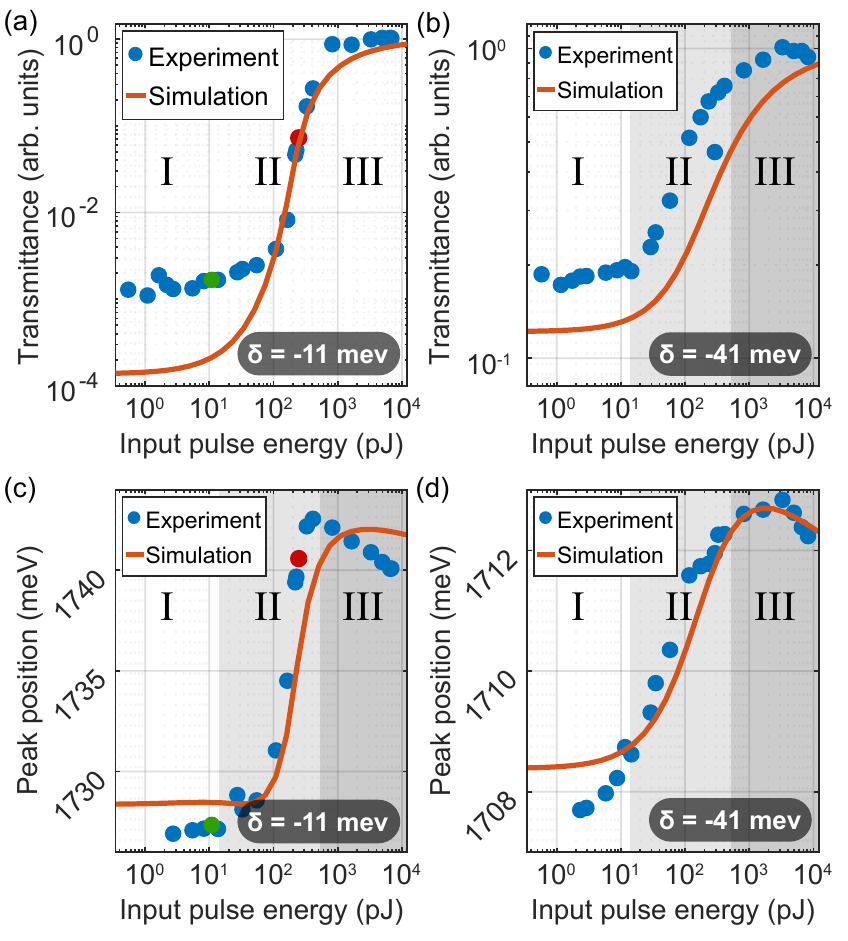}
\caption{\label{fig:3}  Power dependencies of the laser pulse (a,b) total transmittance and (c,d) peak position calculated as a first moment for two negative detunings of (a,c) $\delta_1~=~-11~meV$ and (b,d) $\delta_2~=~-41~meV$. Experimental and simulated results are presented by the circles and  solid lines, respectively. In panels (a,b), the three characteristic regions are clearly visible: low-energy linear (constant transmittance) regime (I), intermediate-energy nonlinear (growing  transmittance) region (II), high-energy nearly linear (constant transmittance) mode. In the latter case, the exciton resonance becomes strongly blueshifted such that the majority of the pulse radiation energy propagates without polaritonic absorption. 
}
\end{figure}

Figure~\ref{fig:2}(a,b) shows the evolution of output pulse spectra after the propagation through the waveguide with the increase of the pulse energy. The data are obtained for two negative detunings of the central frequency of the input pulse from the exciton resonance: $\delta_1 = -11~meV$ and $\delta_2 =~-41~meV$. At the small detuning and low pulse energies, the measured transmitted pulse spectra have a pronounced asymmetric shape due to the proximity of the laser pulse central frequency to the exciton resonance, which leads to an efficient absorption in the high-energy range. With the increase of the pulse power, the shape of the output spectrum for $\delta_1 = -11~meV$ detuning becomes more symmetric, and for both series of the measurements, we observe the blueshift-redshift crossover of the power-dependent peak position. Figure~\ref{fig:2}(c,d) shows the results of the corresponding numerical simulations based on the theoretical model described below. 

Note that for strong pumps, the normalized output pulse spectrum becomes nearly pump-independent in both the experiment and simulations. This is due to the excitonic saturation, which leads to the suppression of both the effective nonlinearity and the absorption. For this reason, the output pulse spectrum at high powers almost corresponds to the spectrum of the input pulse coupled into the system (shown with black curves in Figure~\ref{fig:2}). 

To quantify the output laser pulse after its propagation through the waveguide in each measurement,  
we calculate the area under the spectral curve normalized to the incident power and the first moment of the spectral power distribution in order to determine the overall pulse transmittance (Figure~\ref{fig:3}(a,b)) and the output pulse central frequency (Figure~\ref{fig:3}(c,d)), respectively. Both quantities exhibit strongly nonlinear dependencies on the incident pulse energy, which are defined by the microscopic mechanisms of the nonlinear response in the system, as it will be shown with the use of the theoretical model described below.

\subsection{Model}

To describe the evolution of the laser pulses in the transmittance spectra after their propagation through the waveguide with increasing pulse energy, we used a theoretical model that takes into account the strong exciton-photon coupling, exciton-exciton repulsion within the pulse and the presence of an incoherent exciton reservoir. In particular, we assume that photons can be absorbed creating coherent excitons, which can in their turn annihilate emitting coherent photons. This gives rise to the hybridisation of the photons and excitons and formation of the polariton gap in the dispersion of the elementary excitations. The nonlinear effects in such systems appear mainly because of the the exciton-exciton interactions, which shift the exciton resonance frequency.

For satisfactory description of the experimental data, it is also important to account for the formation of an incoherent excitonic reservoir, where coherent excitons created by the optical pulse can scatter. These incoherent excitons lie far outside the light cone, and thus cannot emit coherent photons. However the reservoir density strongly affects the blueshift of the excitonic line and thus can be seen as a non-instantaneous nonlinearity.

We use the following set of the equations:
\begin{subequations}
\begin{align}
 (\frac{\partial }{\partial t} + v_g \frac{\partial }{\partial x} +\gamma_{ph})A=\iota \frac{ \Omega_R}{2}  \psi +f(t, x)\label{eq:main1}  \\
  \frac{\partial \psi}{\partial t}=  -(\gamma _e+\frac{ \mu }{2}) \psi + \iota \alpha(|\psi|^2+\rho)+\iota  \frac{ \Omega_R}{2} A   \label{eq:main2} \\
   \frac{\partial  \rho }{\partial t} = - \Gamma  \rho + \mu \left |  \psi \right |^2 \label{eq:main3} 
\end{align}
\label{eq:model}
\end{subequations}
The guided photons are described by the slow varying amplitude $A$ of the fundamental mode of the waveguide with group velocity $v_g$. The physical meaning of $A$ and $\psi$ is that the photon and coherent exciton densities are expressed as $|A|^2$ and $|\psi|^2$, correspondingly. We neglect the background waveguide and material dispersions because they are much smaller compared to the dispersion caused by the coupling of the photons with the excitons (see~Figure\ref{fig:1}(b). For the same reason we neglect the bare Kerr nonlinearity of an empty waveguide. The effective linear photonic losses (Joule, scattering on the imperfections, etc.) are accounted for by the dissipation rate $\gamma_{ph}$. The interaction strength between the photons and the excitons is characterized by the Rabi frequency $\Omega_R$.  The system is excited by a coherent optical pulse which is transferred to the photonic guided mode by the coupler described by the driving force $f(x, t)$ in the right-hand side of the equation \ref{eq:main1}.

Coherent excitons are described by the order parameter $\psi$, with  $\gamma_{e}^{-1}$ being their lifetime, and parameter $\mu$ the transfer rate to dark incoherent reservoir. The latter is characterized by the density $\rho$ and  dissipation rate $\Gamma$, which is supposed to be much less compared to $\gamma_e$.


The parameters $v_g$, $\Omega_R$ and  $\gamma_e$  are estimated from the measured polariton dispersion and losses (Figure~\ref{fig:1}(b)) as $\hbar \Omega_R= 36~meV$, $v_g=45~\micron~ps^{-1}$. We neglect the photonic losses (as we work in the waveguide regime) and reservoir lifetime (as reservoir excitons are dark and long living),  $\hbar \gamma_{ph} = 0$, $\hbar \Gamma = 0$. 

The nonlinear parameter $\alpha$ is the proportionality coefficient defining the exciton resonant frequency shift $\Delta \omega$ through the the total density of the excitons $\Delta \omega~=~\alpha \left( |\psi|^2+\rho \right)$. We took the typical value of this coefficient for TMD monolayers from literature  
$\hbar\alpha~=~1~\mu eV \micron^2$~\cite{emmanuele2020highly, kravtsov2020nonlinear}. 

The parameters $\hbar \mu=3 meV$, $\hbar \gamma_e= 9~meV$  are chosen to get the best agreement between the theory and the experiment (Figure~\ref{fig:3}). The pump $f$ is amplitude of the incident light multiplied by the transmission coefficient connecting the intensity of the incoming light to the intensity of the light coupled to the guided mode. We fitted this coefficient by comparing the simulated and experimental data.  Careful optimization of the model parameters confirmed that in our system, the Coulomb-type nonlinearity leading to the exciton resonance blueshift (defined by $\alpha$) dominates over another possible source of nonlinearity, namely the quenching of Rabi splitting originating from phase space filling effects.  This conclusion agrees well with  previously reported both theoretical\cite{PhysRevB.102.115310} and experimental\cite{kravtsov2020nonlinear} studies of the nonlinear response of exciton-polaritons in TMD-based systems.

 \begin{figure}[th]
\includegraphics{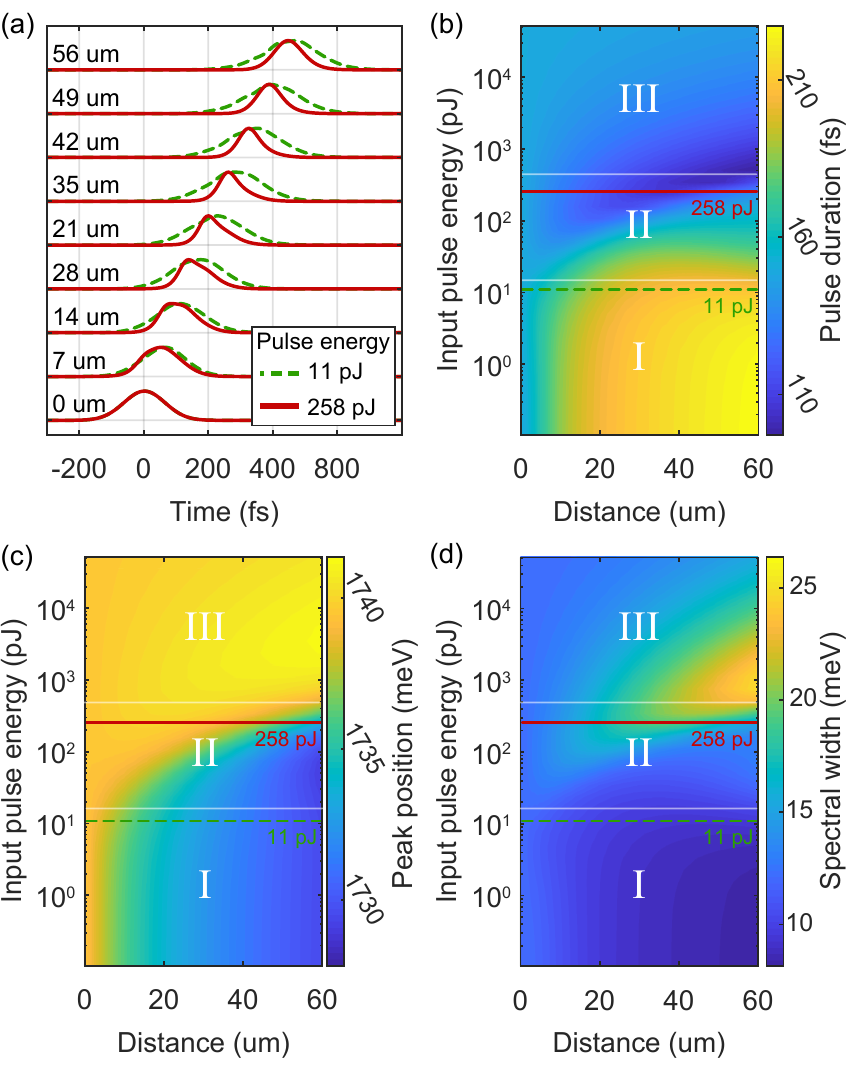}
\caption{\label{fig:4}  (a) Simulated evolution of temporal profile of the pulse propagating along the waveguide for the input pulse energies of 11~pJ and 258~pJ and detuning of -11~meV. In the low-energy regime (11~pJ), the dispersion-driven temporal pulse elongation is observed. On the contrary, for the input pulse energy of 258~pJ, the pulse becomes about $30\%$ shorter and remains stable during the propagation over several tens of microns, which is a signature of a quasi-solitonic behaviour. Simulated (b) duration, (c) peak position, and (d) spectral width of the pulse depending on propagation distance and input pulse energy. Pulse duration and spectral width are determined as a second moment $\mu_2$ of the spectral and temporal power distributions, respectively. For the incident pulse spectrum of a Gaussian shape, the second moment is equal to the variance $\mu_2=\sigma^2$ and related to the full width at half maximum as $FWHM=2\sqrt{2\mu_2\ln{2}}$.
}
\end{figure}

 \section{Discussion}
The theoretical model presented above provides good semi-quantitative description of the experimental data (See Figures~\ref{fig:2} and \ref{fig:3}). 

Let us first analyze the dependence of pulse transmittance on the input pulse energy shown in Figure~\ref{fig:3}(a,b). For both detunings, one can distinguish three characteristic regions. Under both low and high input pulse powers, the system behaves linearly (transmittance is nearly energy-independent). In the first case (Region I), the nonlinearity is feeble, while for the latter one (Region III), only small portion of pulse energy is absorbed before the strong blueshift the exciton resonance, therefore the overall transmitted pulse energy remains almost unaffected. For this reason, we can use the high-energy spectrum as a reference while maximizing light coupling efficiency in the experiment. For intermediate input powers, the system reveals strongly nonlinear response: transmittance increases by several orders of magnitude, central frequency of the output pulse shows complex non-monotonic behaviour, and pronounced spectrum broadening compared to the low- and high-energy regimes appears (see Figure\ref{fig:2}(a and c)). 

First, we acknowledge that at the low input pulse energy, the output spectrum of the pulse is significantly red-shifted compared to the spectrum of the input pulses (black curves in Figure~\ref{fig:2}). The explanation for this is that the contribution of the exciton subsystem to the effective losses is overwhelmingly higher compared to its photonic counterpart. This means that the polaritons experience the highest losses at the exciton resonance frequency. The frequency of the input pulse is lower than the exciton resonance and thus the blue part of the pulse spectrum gets absorbed much stronger than the red part of the spectrum. This explains why the center of the output spectrum is red-shifted in respect to the center of the input one.
 
At intermediate pump energies, the nonlinearity comes in play doing two things. First, it blueshifts the exciton resonance with the increase of input pulse energy. This reduces the difference between the absorption rates for different frequencies, and so one can expect that the redshift of the output pulse becomes lower under higher pump powers. In other words, with the increase of the input pulse energy the central frequency of the output spectrum should undergo \emph{blueshift}.

The second nonlinearity-driven mechanism in the system is the four-wave mixing process that results in the spectral broadening of the pulse directly observed at the intermediate pump powers in the experiment (Figure~\ref{fig:2} (a,c)). Indeed, let us consider the pulse dynamics assuming that frequency-dependant polaritonic losses are not affected by the nonlinear effects (exciton blueshift is disabled) but the spectral broadening takes place. The four-wave mixing increases the width of the spectrum but high frequencies experience the losses dramatically higher and get absorbed much stronger than the lower frequencies. This means that the high-frequency wing of the output spectrum does not grow much -- these frequencies are absorbed. However the lower frequencies, including the ones generated by the four-wave mixing survive and the lower-frequency wing of the spectrum becomes longer. This means that the central frequency of the pulse should redshift. Therefore, if the process of spectrum broadening is efficient and dominates over the nonlinear modification of the effective losses, with the increase of the input pulse energy the central frequency of the output pulse decreases, i.e. undergoes \emph{redshift}.

If we perform numerical simulation with account for these two competing nonlinear mechanisms with no reservoir enabled (see Supplementary Material, Fig S1 for details), we see that the behaviour of the central frequency is not monotonic. Only at high powers the exciton blueshift reduces the losses for higher frequencies strongly enough to suppress the initial redshift. Meanwhile, in the experiment, no redshift at intermediate pump pulse energies is observed (see Figure\ref{fig:2}(a,b), Region II).

 In order to explain the experimentally observed behavior of the central frequency of the output pulse spectrum we can surmise that the four-wave mixing in the experiment is less efficient compared to the model without a reservoir. So what we need is to suggest such a modification of the model that would suppress the efficiency of the new frequency generation on the polariton density. This can be done by an introduction of the non-instantaneous nonlinearity, assuming that the coherent excitons can get scattered  into the incoherent ones. The incoherent excitons do not take part in the four-mixing process spreading the spectrum but still contribute to the exciton blueshift and thus nonlinear blueshift of polaritonic  losses. Usage of this model (eqn.~(\ref{eq:model})) allows us to obtain good semi-quantitative agreement with the experiment. It is worth mentioning that there may be other physical mechanisms (for example generation of electrons-holes plasma) leading to a similar effect. However, in the studied system the best agreement was obtained for the model accounting for the reservoir, i.e. generation of the incoherent excitons. 
 
 Let us now mention the possible reasons for the discrepancy between the experiment and the simulations. Note that in the system under study, losses depend on frequency that strongly that the attenuation varies by at least three orders of magnitude for the given propagation distance within the spectral width of the initial pulse. This leads to the fact that for reproducing the attenuation curve quantitatively, we need to know the spectrum of the initial pulse with the precision much exceeding experimentally achievable values.  
 

The experimental data on the power dependent reshaping of the spectrum of the output signal allows us to reconstruct the spatio-temporal dynamics of the pulse. Figure~\ref{fig:4} shows the simulation results for pulse propagation in the polariton waveguide. Under low input energies (Region I), pulse undergoes polaritonic absorption leading to the redshift of its spectrum and strong temporal broadening (see also green dashed curve in panel (a)). For high input pulse energies (Region III), pulse parameters remain nearly unchanged due to the strong exciton blueshift and thus weakening of the nonlinear processes. The intermediate regime (Region~II) is illustrated in Figure~\ref{fig:4}(a). While propagating, the pulse first gets compressed, then its duration and spectral width remain nearly constant for approximately 30 microns. Such a behaviour clearly indicates a quasi-stationary solitonic regime of pulse propagation, which is realized in the studied system.

\section{Conclusion}
We have shown that strong light-matter coupling in a TMD-based polaritonic waveguide gives rise to pronounced nonlinear effects  in propagation of ultrashort optical pulses. We observe power dependent suppression of the losses at large pulse energies and blueshift-redshift crossover of the pulse central frequency. These features are well-reproduced by a theoretical model which accounts for the excitonic reservoir and Coulomb interaction between the excitons. The spectral broadening experimentally observed at intermediate pulse powers is a manifestation of a new frequency generation process. Our modelling confirms that it is accompanied by the temportal shortening of the pulse, which is the fingerprint of the quasi-stationary solitonic regime.

\section{Acknowledgments}
The work was partially funded by Russian Science Foundation, project 21-72-10100 and Priority 2030 Federal Academic Leadership Program. I.A.S. acknowledge support from the Icelandic Research Fund (Project "Hybrid polaritonics").

\section*{Methods}
\subsection*{Sample fabrication}
To realize the proposed polaritonic system, we chose an unpatterned Ta$_2$O$_5$ waveguide on a SiO$_2$ sublayer. Ta$_2$O$_5$ layers of 90 nm thickness were deposited on commercial SiO$_2$/Si substrates via e-beam assisted ionbeam sputtering. Grating couplers were fabricated by patterning the upper dielectric layer via a combination of electron-beam lithography and plasma etching to yield the following geometric parameters: pitch of 430 nm, groove width of 200 nm, and depth of 30 nm.

After preliminary characterization of the waveguide transmission, the WSe$_2$ monolayer, which was fabricated by mechanical exfoliation from a commercial bulk crystal from HQ Graphene, was transferred onto the waveguide surface between the grating couplers. The monolayer length was 50~$\mu m$ along the pulse propagation direction.
\\
\subsection*{Optical measurements}
Optical measurements were performed using the back focal (Fourier) plane imaging in our custom-built experimental setup, which allows for direct extraction of the dispersion diagrams in frequency-momentum space (see Figure~\ref{fig:1}(b)). The laser pulses and white light (polarised along the grating that corresponds to transverse-electric (TE) field mode) were focused onto the input grating coupler using an objective lens. Light decoupled from the waveguide by the output grating was collected by the same microscope objective (Mitutoyo Objective 50x/0.65) and spatially filtered. As a result, only radiation from the output grating was detected. The angle-resolved signal was measured by projecting the Fourier plane of the objective onto the entrance slit of an imaging spectrometer (Princeton SP 2550) and detecting the light with a CCD detector (PyLoN 400BR eXcelon). 
The sample was mounted in an ultra-low-vibration closed-cycle helium cryostat (Advanced Research Systems, DMX-20-OM) and maintained at 7~K.


\bibliography{main}
\bibliographystyle{naturemag}

\end{document}